# Finite bias dependent evolution of superconductor-insulator transition and Zero Bias Conductance in boron doped nanodiamond films

Davie Mtsuko, Christopher Coleman, and Somnath Bhattacharyya*

*Nano-Scale Transport Physics Laboratory, School of Physics and DST/NRF Centre of Excellence in Strong materials, University of the Witwatersrand, Johannesburg 2050, South Africa*

We report on transport features in heavily boron doped nanocrystalline diamond (BNCD) films which are not seen in conventional (s-wave) granular superconductors. Observations include an anomalous resistance peak near to the superconducting transition temperature as well as a strong zero bias conductance peak in the current-voltage spectra. The effect of finite bias current on the evolution of the resistance peak is systematically investigated in this system. The shape of the resistance-temperature curves near the critical temperature is seen to be strongly influenced by both magnetic field and bias current. As the bias current is lowered the resistance peak becomes more pronounced whereas when the magnetic field is varied the peak shifts towards lower temperatures, the resistance upturn shows a quadratic temperature dependence as expected for a Kondo transition. We find that a number of transport features such as resistance peak height, zero bias conduction peak height and width as well as magnetoresistance peaks scale according to a power law dependence. We interpret these features as a result of a charge-Kondo effect where hole dopants act as degenerate Kondo impurities by opening additional pseudo-spin scattering channels.

## I. INTRODUCTION

Pristine, conventional and bulk 3D superconductors usually exhibit a metal-superconductor phase transition with a feature-less zero resistance phase below a global critical temperature ($T_c$) [1]. On the other hand 1D and 2D superconductors often demonstrate complex phase transitions that include an insulating phase with an anomalous resistivity peak in the vicinity of $T_c$ [1]. It is well known that heavily boron doped diamond (HBDD) (films) exhibits a superconducting phase transition at low temperature with a transition width-critical temperature ratio ($\Delta T/T_c$) that varies with grain size [2]. Further, it has recently been proposed that granular 3D heavily boron doped diamond (film) undergoes a metal-bosonic insulator (BI)-superconductor transition resulting in a giant anomalous resistivity peak (referred to as a bosonic insulator phase) in the vicinity of $T_c$ [1]. In previous studies, Zhang *et. al*. explained the formation of this peak as the generation of bosonic islands that absorb fermions. The peak was considered unconventional since the ratio of the peak resistivity to the normal metal phase threshold resistivity (up to ~$10^3$%) was found to be much greater than the thermo-resistivity <10% in weak localization phenomena [1]. The sharp upturns in the resistivity leading to the so called BI peak does however resemble resistance transitions observed in materials that exhibit the Kondo effect [3]. The observation of the Kondo effect is usually associated with the presence of magnetic impurities as discussed in reports of Kondo effect in materials such as $(La_{1-x}Ce_x)Al_2$ [3]. However, strictly speaking, any degenerate two level system that is coupled to a continuum of carriers can potentially demonstrate a Kondo effect, a notable example of this is the charge analogue of the spin-Kondo effect whereby a Fermi sea couples to impurities that have degenerate charge states. Such charge-Kondo effects have recently been investigated in strongly Coulomb repulsive quantum dots as well as non-superconducting single electron devices [4,5]. Another active and perhaps more relevant area of researched related to the charge-Kondo effect is in the valence-skipping elements with attractive onsite interactions (called negative U molecules). In such materials the degenerate valence states of the impurities open up additional scattering paths that lead to a Kondo effect. Such phenomena have already been observed in Thallium (same group as Boron) doped PbTe, where the Tl dopants induce a superconducting phase and exist as either a $Tl^{3+}$ or a $Tl^{+}$ valence states, this system has demonstrated resistance upturns (re-entrant effects) at low temperatures similar to our material of interest, i.e. HBDD. Additionally the boron acceptor state (i.e. hole bound to boron impurity) in doped diamond is known to be four fold degenerate and has demonstrated spontaneous symmetry breaking related to spin-orbit splitting at low temperatures [6,7]. This highlights some interesting possibilities in relating the anomalous transport features observed in HBDD (films) to a charge-Kondo effect similar to the negative U behaviour demonstrated by Tl doped PbTe. Additionally it should be noted that the HBDD has a complex microstructure comprising of nanoscale diamond grains separated by graphitic $sp^2$ carbon similar to disordered graphene, which is in itself an interesting phase of carbon believed to allow anisotropic multi-channel Kondo effects [8]. It is therefore interesting to investigate the effect of charging due to low dimensionality and how such a charge-Kondo phenomenon relates to previous reports on charging effects in granular superconductors [9-11]. Similar transport features have been observed in unconventional superconductors where spin impurity correlations or non-s wave symmetry order parameters have been established [12-22].

In this work we systematically study the finite bias current and magnetic field dependence of the superconductor-insulator transition in BNCD films in high bias current regime as well as in a bias regime well below the critical current ($I_c$). We present the observation of re-entrant peak features below the onset of superconductivity which become more pronounced as the current is decreased. We argue that this behaviour emanates from a competition between Bardeen-Cooper-Schrieffer (BCS) superconductivity pairing and scattering effects due to inter-grain or intra-grain tunnelling processes involving bound hole states influenced by the Coulomb repulsion ($U$). Further, we show that the peak resistance follows a power law temperature dependent scaling and also ascertain that what has been referred to as a bosonic insulator peak is actually a signature of Kondo behaviour in BNCD films. One of the most important features reported here is the differential conductance peak measured at zero bias, this anomaly has shown a strong magnetic field and temperature dependence, and has before been related to Kondo scattering and Andreev Bound States (ABS) in superconducting tunnel junctions, thus supporting our interpretation of a charge-Kondo effect. These analysis are supported by the magnetoresistance features, particularly the transition from a positive to negative resistance phase. Although resistive transitions have before been reported for granular superconducting systems in terms of a charge-vortex duality [9], the anomalous peak in the $R(T)$ and magnetoresistance features as well as the zero bias conductance peak is not a general feature of granular superconducting systems. The set of transport properties suggest the unconventional nature of superconductivity in BNCD films and indicate non-s type order parameter.

## II. EXPERIMENTAL DETAILS

**Table I.** Transport parameters for superconducting nanodiamond films such as critical temperature at zero field ($T_{c\text{-mid}}$, mid implies the value of $T_c$ that is evaluated in the mid of the SC transition where the resistance falls within 50% of the normal state resistance value, critical magnetic field ($B_c$) critical current ($I_c$), superconducting gap $\Delta$, universal scaling exponent, $\beta$ (slope of log $R$ vs. log $I$ data), grain size ($D$) and Kondo temperature ($T_K$), see text for details of calculations. Sample B4 and B2.5 are studied here due to their intermediate grain size (60-70 nm).

| Sample | $T_c$ (K) | $B_c$ (T) | $I_c$ (μA) | Δ (meV) | $\beta$ | $D$ (nm) | $T_K$ (K) |
|---|---|---|---|---|---|---|---|
| B1 | 3.6 | 2.5 | >800 | 0.56 | ~0.9 | 100 | ~3.5 |
| B2.5 | 1.2 | 0.9 | 25 | 0.18 | 0.94 | 60 | 1.5 |
| B4 | 1.8 | 1.0 | 40 | 0.27 | 0.97 | 70 | 0.8 |
| B5 | 2.0 | 1.1 | 20 | 0.30 | ~0.9 | 30 | ~0.5 |

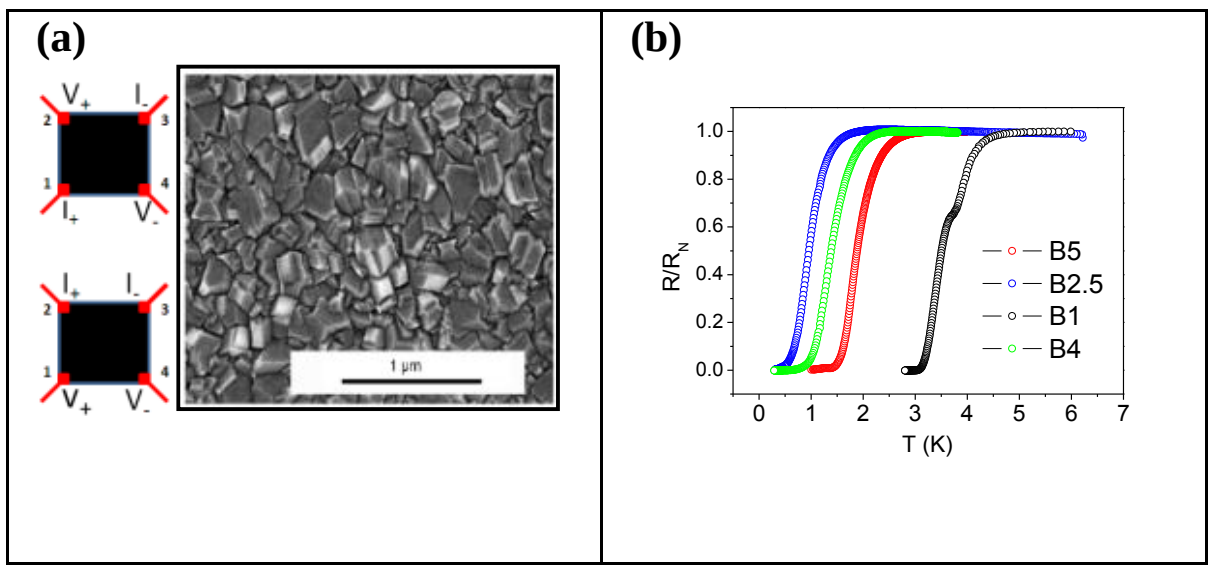

**FIG. 1.** Morphology and basic superconductivity transport measurements of boron doped nanocrystalline diamond films (BNCD): (a) Scanning electron microscope (SEM) image of a typical 400 nm thick BNCD film. Inset: van der Pauw transport measurement configuration. (b) Variation superconducting transition between samples plotted as normalized resistance, the transition ot zero resistance regime is clearly observed.

The BNCD samples investigated in this work were synthesised using microwave plasma enhanced CVD. The diamond samples were grown with 1%, 2.5%, 4% and 5% $CH_4$ in $H_2$ (henceforth named B1, B2.5, B4 and B5) with trimethylborane (TMB) as a dopant precursor on quartz substrates. The ratio of TMB:$CH_4$ was ~4000 ppm. Scanning electron microscopy images show that boron doped samples have a granular surface morphology with grain sizes that vary from 20-30 nm for sample B5 and 100 nm for sample B1 (See Fig. 1(a)) [2]. Resistance measurements were performed in a four-probe van der Pauw geometry. Longitudinal resistance $R_{XX}$ was measured by applying current on one edge and measuring voltage drop on the opposite end (See Fig. 1(a) inset). Transverse resistance ($R_{XY}$) was similarly measured by applying current across two diagonally opposite corners of a 5 mm × 5 mm chip and measuring the voltage across the other two diagonally opposite vertices (See Fig. 1(a), inset). The magnetic field was applied perpendicular to the diamond film. The temperature stability was better than 10% at 300 mK. The existence of superconductivity is confirmed by measuring resistance as a function of temperature as well as at various magnetic fields and fixed bias currents. As shown in figure 1 (b) it was found that all the samples undergo a metal-superconductor transition with critical temperatures in the range 1.2 K - 3.6 K. Similarly, voltage-current characteristics and the differential conductance were studied for these samples at different experimental conditions. We have also investigated the magnetoresistance of the samples at various bias strength and temperature. Although all samples demonstrated the features reported here, we have chosen to display data from two samples in particular, namely B2.5 and B4 because of their intermediate grain size (60-70 nm) which give a good representation of the set of transport phenomena that are the topic of this paper. Table I shows a list of the various measured transport parameters of BNCD samples studied in this work. The models and assumptions used to calculate the additional parameters such as the Kondo temperature are discussed in the respective sections.

## III. RESULTS AND DISCUSSION

In section III A, we first present the overall temperature, magnetic field and bias current strength dependence of the resistance. This allows us to establish various phase diagrams related to the observation of the resistive peak. The upturn leading to the resistance peak as well as its scaling properties are analysed in light of a Kondo effect allowing for the extraction of the Kondo temperatures for the respective samples. Section III B, is concerned with the *I-V* characteristics of the samples. We show the power law dependence of voltage on current and establish a BKT transition and hence two dimensional character of the granular films. A pronounced zero bias conductance peak (ZBCP) is observed in the differential conductance which is analysed and related to Kondo impurity state and ABS. Finally, section III C is focused on the observation of temperature dependent magnetoresistance, where a transition from a weak anti-localization (WAL) regime at low temperatures shows a transition to a weak localization (WL) regime near to the derived Kondo temperature. These features are discussed in light of pseudo-spin scattering that takes place between non-equivalent valleys which we argue arise due to graphitic (graphene-like) grain boundary regions.

### A. Temperature dependent resistance and re-entrant behaviour:

As previously mentioned, the upturn and peak feature in the temperature dependent resistance has before been observed in boron doped diamond [1] and explained qualitatively as the formation of a Bosonic insulating phase. This phase is supposedly formed by localized Cooper pair islands that grow as temperature is lower until a critical density is reached, decreasing the temperature further results in global superconductivity. The same resistance peak has been identified in our samples as can be seen in figure 2 (a) which shows the re-entrant peak feature at the foot of the temperature dependent transition at zero applied magnetic field, also shown in figure 2 (a) is the suppression of this peak as the field strength is increased.

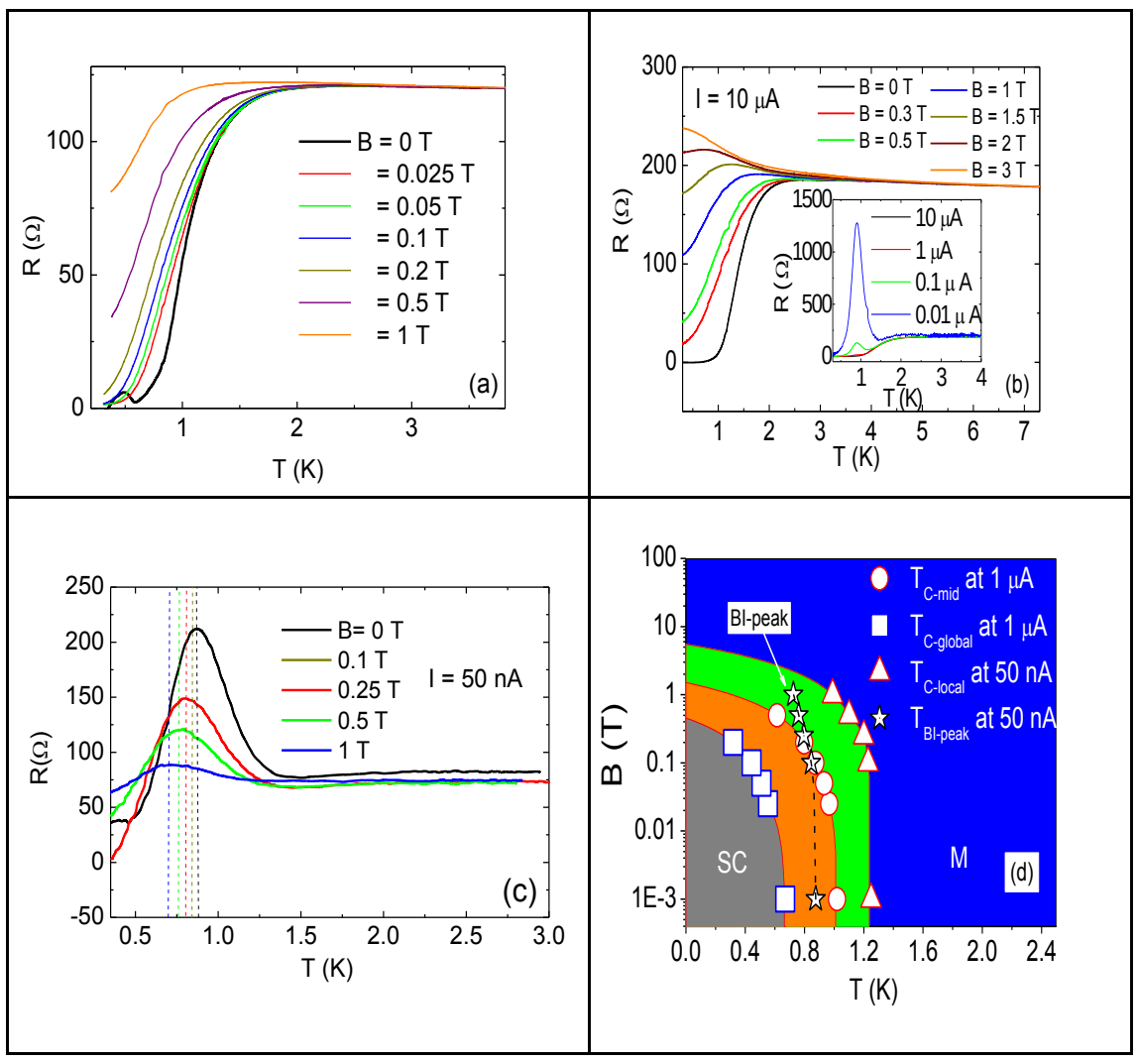

**FIG. 2.** (a) Evolution of the field induced superconductor-insulator transition for sample B2.5, the presence of the re-entrant resistance peak is observed clearly when no magnetic field is applied (b) Variation of resistance peak position with applied field for sample B2.5 at a bias current of 1 μA. (c) Evolution of the insulator-superconductor transition with magnetic field at a low bias current (10 nA) for sample B4. The bosonic insulator peak intensity can be greatly enhanced by decreasing the measurement current as shown in the inset. (d) A plot of $T_{c\text{-global}}$, $T_{c\text{-mid}}$, $T_{c\text{-local}}$ and $T_{BI\text{-peak}}$ as a function of magnetic field showing the various phases for sample B2.5 (see text for details).

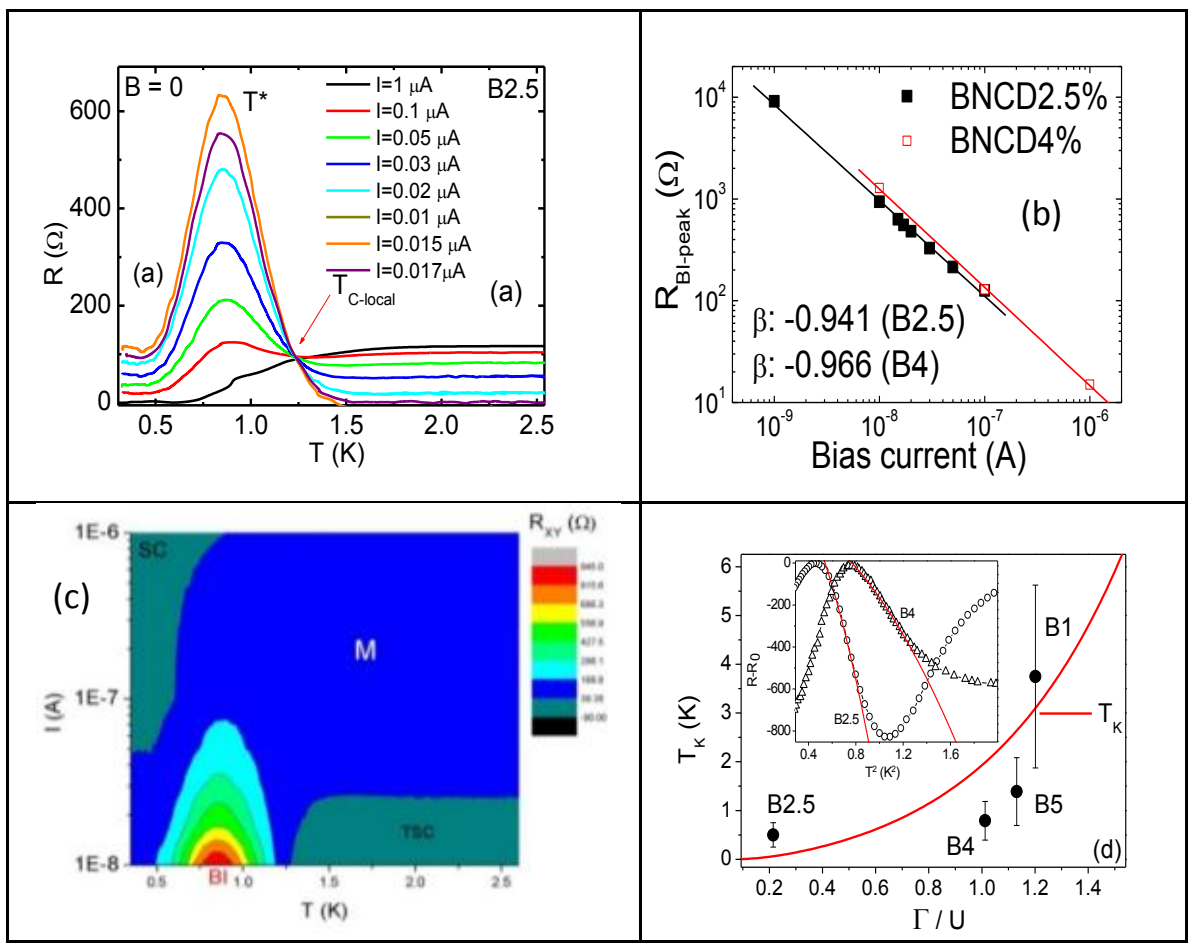

**FIG. 3**. (a) Evolution of the resistive peak as measurement current is changed, the conditions that favour the insulating phase is that of low bias. (b) The resistance peaks of the samples B4 and B2.5 show an exponential decrease in amplitude with increasing magnetic field strength whereas the peak height is found to scale according to a power law with an apparent universal exponent of $\beta = 1$ as shown in the inset of Fig 2(b). (c). The phase diagram derived from temperature dependent resistive peak at various applied bias current strength. (d) The derived Kondo temperatures of all four samples as a function of ratio of tunnelling coupling to onsite charging energy (Eqn. 3). Inset: The respective Kondo temperatures are obtained from fitting $T^2$ dependence of the upturn in the resistance.

It was found that this resistive peak showed a great degree of tuneability and that the measurement conditions, particularly the bias current and applied magnetic field could greatly enhance or suppress the peak. Figure 2 (b) shows the temperature dependent resistance at different applied fields, at 10 μA no peak is observed and the field drives the system to the insulating state, however as shown in the inset of figure 2 (b), the peak appears and increases in height as the bias current is lowered. Thus by lowering the measurement current it is possible to tune the respective samples to a regime where the resistance peak is pronounced and can be qualitatively investigated, this is demonstrated for sample B2.5 in figure 2 (c) where upon lowering the current to 50 nA the effect of the magnetic field can be clearly seen. As reported before [1] the peak shifts to lower temperatures as field is increased, the peak height is also influenced by the magnetic field and decreases exponentially as field is increased.

In order to correctly capture the nature of the resistive peak, the phase diagram given in figure 2 (d) was constructed by tuning the system between two regimes (one where the peak is pronounced (50 nA) and the other where the peak is suppressed (1μA)) and measuring the temperature and field dependent resistance. At higher current the phase diagram has only two relevant critical points, the mean field critical temperature ($T_{c\text{-mid}}$ defined as the mid-point of the transition, indicated as circles in the phase diagram), and the point at which the system is superconducting ($T_{c\text{-global}}$, squares). In the low current regime, the resistive peak upturn ($T_{c\text{-local}}$) and the resistive peak maximum ($T_{BI\text{-peak}}$) are the critical points of interest. The phase diagram given in figure 2 (d) is a comparison of the trend of these four critical points (i.e. below and above resistance peak conditions) and as can be seen in the low bias regime, the peak resistance upturn always occurs at higher temperatures than the mean field critical temperature, an indication that electrons correlations in the fluctuation regime are related to the origin of this phenomenon. Another observation of interest is the deviation of the $T_{BI\text{-peak}}$ (data points given by stars) magnetic field dependence from the other critical points field dependence. This trend cannot be accounted for by the Ginzburg-Landau theory, which does however neatly fit the other three critical points temperature/critical field relationship and allows for the extraction of the Ginzburg-Landau coherence length:

$$\xi_{GL} = [\Phi_0/2\pi B(T\sim 0\ K/2)]^{1/2}, \qquad (1)$$

where $\Phi_0 = h/2e$ [1]. The coherence length for sample B2.5 was found to be $\xi_{GL}$~ 27 nm which is smaller than the grain size of the respective sample ($D$ ~ 60 nm) and is higher than that reported for 3D heavily doped diamond ($\xi$ ~5.5 nm) and nearly twice that of monocrystalline diamond ($\xi$ ~15 nm) [21] indicating that although granular in nature the samples are of high quality and do not suffer from high defect density.

As the effect of the bias current is strongly related to the observation of the resistance peak, a thorough investigation of the peak as a function of current has been undertaken. As can be seen in figure 3 (a), the transverse ($R_{XY}$ or Hall component) of the resistance also shows the pronounced resistance peak features which increase in intensity greatly as bias current is decreased. In fact the peak is found to scale to a power law dependence as a function of the measurement current. The power law behaviour for samples B2.5 and B4 have been shown in figure 3 (b), the critical exponent describing the scaling is found to be approximately $\beta$ ~ -1. This value is found throughout the samples and is thus believed to be a universal value for this system.

The phase diagram given in figure 3 (c) is constructed by measuring the temperature dependence of the resistance for a large range of measurement current strength, as can be observed the large increase in resistive peak height is only observed below approximately 100 nA and in zero applied field exists around the fluctuation regime of the superconducting transition. The enhancement of the peak at low current, i.e. more resistive regime is believed to be a result of increased electron-electron interactions due to enhanced Coulomb repulsion. Taking into account the granular structure of the material, charging effects as those observed in Josephson junction arrays are expect to play a role in the transport features, it is then natural to interpret the re-entrant features within a Kondo model resulting from enhanced scattering in a resistive regime due to the interplay of $U$ and $\Gamma$ (the coupling of grains).

The interplay of superconductivity and the Kondo effect has been predicted to exist in superconductors with magnetic impurities [22,23] and has already been observed in a superconducting ($La_{1-x}Ce_x$)Al alloy [5]. Exotic Kondo features can also be observed due to non-magnetic impurities as suggested by studies conducted on PbTe which superconducts when doped with Tl and has demonstrated fluctuation charge-Kondo effects [20]. In such charge-Kondo systems negative-$U$ Hubbard interaction between hole carriers have been used to explain the charge pairing mechanism behind superconductivity, the Kondo effect was argued to exist due to scattering from degenerate charge states of the Thallium dopants ($Tl^{3+}$ and $Tl^{+}$). One well known degenerate state that might be related to the Charge-Kondo effect here is the boron acceptor, i.e. bound hole state, which is known to be four-fold degenerate. In HBDD these acceptor states are responsible for the conduction due to overlap and hybridization of the hole states forming an impurity band (consisting of the four-fold degenerate modes) near to the top of the intrinsic valence band of the diamond. Thus in the low current regime, where the re-entrant resistance peaks are most dominant, it is possible to consider a charge-Kondo effect resulting from scattering related to these degenerate hole modes. Following the logic presented in reference 20 (i.e. the investigation of charge-Kondo effect in Tl doped PbTe) the upturn leading to the resistance peak is analysed in light of a charge-Kondo effect. As shown in the inset of figure 3 (d), the increasing resistance shows a $T^2$ dependence as expected for a Kondo scattering regime. It is possible to extract a Kondo temperature by fitting the data shown in the inset of figure 3 (d) with

$$R - R_0 = R_{min}\left[1 - \left(\frac{T}{T_k}\right)^2\right], \qquad (2)$$

where $R_0$ and $R_{min}$ are the peak and minimum resistance values respectively. This analysis is performed on all four samples and respective Kondo temperatures are given in table 1. To verify the suitability of a Kondo effect and test the validity of the Kondo temperature determined from the upturn in resistance we compare the results to the Wilson numerical renormalization group theory. This model is valid for crossovers from singlet to spin-triplet pairing and has before been used for re-entrant superconductivity phases in a wide range of systems including carbon nanotube devices, accordingly the Kondo temperature is known to follow the relation [24]:

$$T_K = \sqrt{\frac{U\Gamma}{2}}\exp\left[\frac{-\pi U}{8\Gamma} + \frac{\pi\Gamma}{U}\right] \qquad (3)$$

In order to make a meaningful comparison we rely on the fact that granular superconducting films are widely known to exhibit transport features related to Josephson junction arrays, in such systems the respective array elements have a charging energy and are coupled through a Josephson coupling term [25,26], as a first approximation these are analogous to the onsite energy and coupling parameter used in the Kondo model. The charging energy is assumed to be constant for all samples as it depends only on the capacitance of the material ($E_C = e^2/2C$ and C ~ 67 pF established via atomic force microscopy studies from a typical BNCD film [27]) and the Coupling energy ($E_J$) is determined by fitting the resistance data with the following formula derived by Tinkham [28] for granular films and arrays:

$$\frac{R(T)}{R_N} = ce^{-b/\sqrt{\tau - \tau_v}} \qquad (4)$$

Where $c$ and $b$ are constants and $\tau$ is the normalized temperature related to the coupling energy through: $\tau = \frac{k_B T}{E_J}$. The $\tau_v$ term is the normalized BKT transition temperature determined from the *I-V* analysis shown in the next section. Thus once the charging and Josephson coupling energies are known, it is possible to relate the Kondo temperature derived from the upturn in the resistance data to the theoretical Kondo relationship given by Eqn. 3, this is shown in figure 3 (d). The data points derived from the experimental data follow the theoretical trend surprisingly well considering the assumptions mentioned above, and although there is some deviation the overall trend follows what is predicted for a Kondo system.

### B. Current-Voltage analysis:

It is well known that 2D superconducting systems under influence of magnetic flux pinning can exhibit the BKT transition due to vortex-antivortex pairing. However, BKT transition has not yet been firmly established in nanodiamond films. Using the conventional scaling analysis and determining the temperature dependence of the critical exponent controlling the power law characteristics of the current voltage response it is possible to identify a BKT transition temperature, shown in Fig. 4(a). According to theory this power law scaling in the current voltage data is related to the so called jump in superfluid stiffness and a transition from unbound to bound vortex/charge states occur when the critical exponent is equal to 3, this is shown in the inset of figure 4(a) for samples B4 and B2.5. Sample B4 shows a clear jump of the critical exponent value past 3 at approximately 0.6 K, for sample B2.5 however, the upturn is observed but no clear BKT transition in our experimentally accessible temperature range. The observation of the BKT transition in these films is somewhat of a surprise as the phenomena is generally restricted to two dimensional materials such as ultra-thin films and fabricated Josephson junction arrays. We explain such observation as resulting from the percolation conduction channels that exist through the grain boundaries, as has been reported for ultra-nanocrystalline diamond films [29,30,31]. Such boundary conduction channels have been reported to be formed from a $sp^2$ hybridized graphitic carbon similar to disordered graphene [32], additionally there have also been studies suggesting structurally arranged boron of exist within the boundaries leading to interfacial superconductivity [33]. As explained in the previous section there is a strong possibility for a charge Kondo effects in this system, this leads to interesting possibility of a charge-BKT instead of conventional vortex-BKT transition occurring here. In addition to the BKT transition and Kondo features described above we have observed a zero-bias conductance (ZBC) peak shown in Fig 4(b).

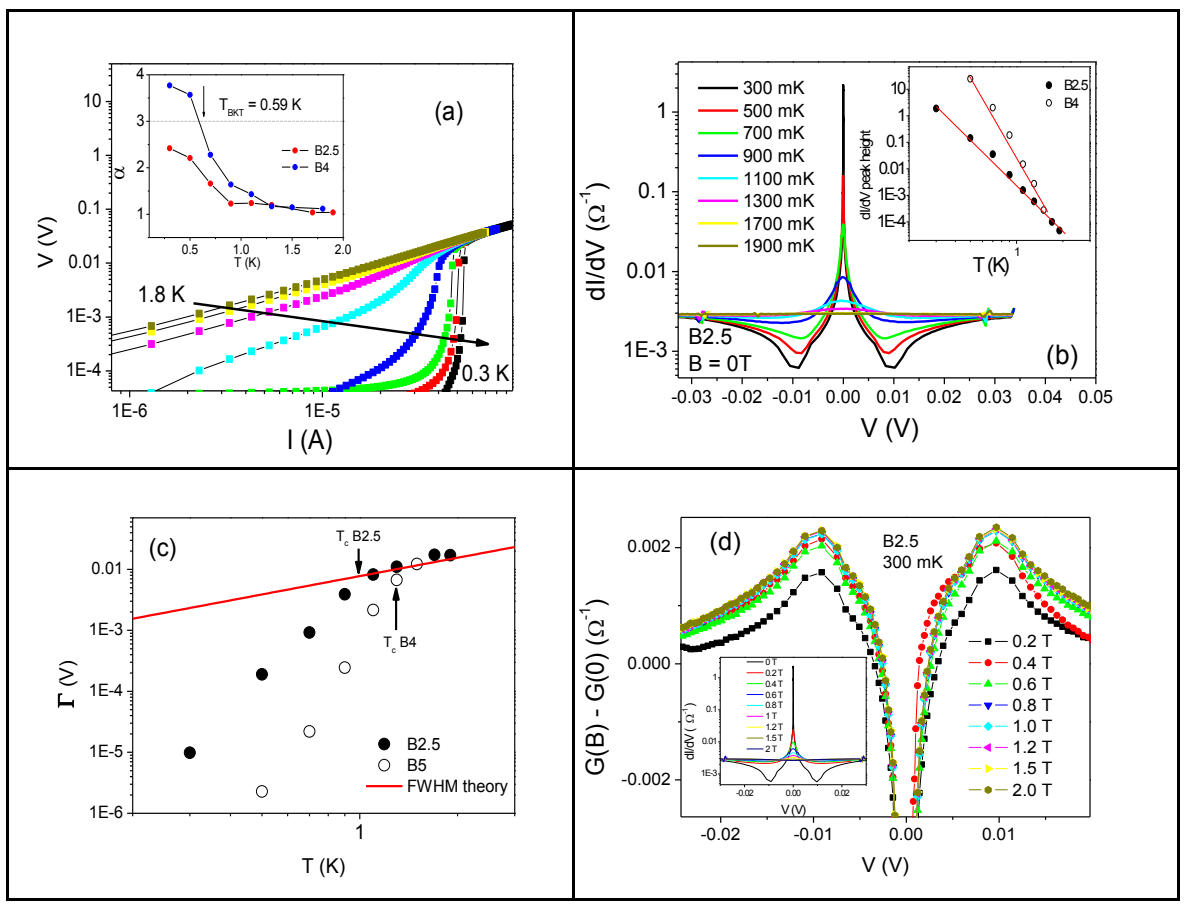

**FIG.4**.(a) Current –voltage response plotted in double log scale, the BKT transition is established by following the temperature dependence of the power law depended of the *I-V* curves, the temperature dependence of the α-parameter for samples B2.5 and B5 are shown in the inset of Fig 4 (a). (b) A pronounced zero bias conductance peak is observed in the differential conductance plots of all samples, demonstrated in Fig 4(b) is such feature for sample B2.5, as shown in the accompanying inset the peak height of samples B4 and B2.5 scale according to a power law, like what was observed for the resistance peak shown in Fig 2(b), inset. (c) The FWHM of the ZBCP for samples B2.5 and B4 show similar temperature dependence, above $T_c$ the FWHM follows the expected fermi liquid behaviour frequently used for Kondo ZBCP however at temperatures below $T_c$ the FWHM decreases with temperature according to a power law scaling. (d) The split peak structure after subtraction of the zero field ZBCP, this allows for evaluation of the split peak structure which in our case do show the expected linear shifting with applied magnetic field. The inset shows the field dependent ZBCP that the main graph was obtained from.

This feature is commonly reported to be a signature of Kondo resonance. ZBCPs have been observed in both spin and pseudo-spin correlated systems, however the analysis for the latter is not as well-known and is currently the focus of intense investigation in the field of quantum dot and mesoscopic superconductor channel research [34,35,36]. In order to further investigate the plausibility of a charge-Kondo effect we analyse the ZBCP in light of conventional Kondo theory. The two main features of ZBCPs that are routinely analysed are the temperature dependent peak height and full width at half maximum (FWHM). For a spin/pseudo-spin Kondo effect the differential conductance peak height is expected to have a logarithmic temperature dependence. This is however not observed in our samples and as shown in the inset of figure 4 (b) the peak height instead follows a power law scaling. To our knowledge this has not yet been observed in ZBCPs and there does not exist any theoretical model to fit this trend. However, when analysing the FWHM temperature dependence, we can clearly identify two different scaling regimes dictated by the critical temperature of the samples. This is shown in figure 4 (c), where the FWHM is observed to closely follow the expected fermi liquid temperature dependence at temperatures above $T_c$ but crosses over to a power law dependence at temperatures below $T_c$. The FWHM in the Kondo regime is given by the following equation:

$$\Gamma = \frac{1}{2}\sqrt{(\alpha K_b T)^2 + (2K_b T_K)^2}\,, \qquad (5)$$

Where $K_b$ is the Boltzmann constant and $T_K$ is the Kondo temperature obtained from fitting the $T^2$ dependence of the resistive peak. Such features have to our knowledge not yet been reported for the ZBCP of superconducting systems, these transport features do however imply that the formation of the peak above $T_c$ may be the result of pre-formed Cooper pairs above $T_c$ and hence a pseudo-gap. A ZBCP has before been observed in high temperature superconductors where such transport features were described as resulting from either spin correlated tunnelling due to magnetic impurities in the tunnel barrier [37,38] or to the anisotropic d-wave order parameter which are essentially the result of the formation of Andreev bound surface states [39,40,41]. The observation here therefore raises some fundamental questions as to the nature of superconductivity in boron doped diamond as the ZBCP relating to ABS is a feature of high $T_c$ materials and is related to the layered or 2D structure of such systems. In fact the observation of the ZBCP is generally used as a means of establishing d-wave, or more specifically non s-wave nature of superconductors. Following convention [37] the ZBCP can be re-plotted as the difference between field dependent isotherms and the zero field sweep ($G(B)–G(0)$), this allows for the subtraction of the steep mid peak and highlights the split peak features which, for spin correlated system are expected to follow a linear Zeeman splitting with increasing field strength [41].

Deviation from the linear splitting has before been used as verification of d-wave order parameter and is expected to result from in-plane mid gap states and not Zeeman splitting of ground states of magnetic impurities [39]. The system at present does not exhibit the expected Zeeman shifting when the field is increased, this may be interpreted as an absence of magnetic impurities supporting the idea of a charge-Kondo effect where holes are related to the scattering events instead of impurity spin. Even more appealing is the possible formation of the ABS (as in the case of d-wave order parameter materials) probably existing within the grain surface or boundary where the boron and hence holes are expected to reside in abundance. The formation of an ABS at the boron acceptor site is expected to have huge implication on the symmetries that define the system. As mentioned before the acceptor state is four-fold degenerate, but however experiences spontaneous symmetry breaking and removal of degeneracy due to spin-orbit splitting arising from a static Jahn-Teller effect. A possible explanation of the observed charge-Kondo effect can then be related to scattering from ABS. When an electron binds to the hole of the acceptor state and forms the ABS this additional electron is believed to act in a way that restores the symmetry of the system, this will restore the degeneracy of the hole energy levels, once this is achieved a charge-Kondo effect is possible due to scattering channels opening up due to the degeneracy of the bound hole (ABS) state being restored.

**C. Low bias current magnetoresistance:**

The ABS is usually associated with surface states due to interfacial contact between superconductor and insulator or within the ab-plane of high $T_c$ d-wave ceramics. The ABS is then a possible indication of two-dimensionality, which is complimented by the observation of the BKT transition reported in the previous section. As stated before, there exists strong evidence for the occurrence of graphitic multi-layered $sp^2$ hybridized carbon along the grain boundaries of the nanocrystalline diamond films [40] and that these grain boundaries provide the channels allowing for a percolation conductance. The occurrence of low dimensional carbon boundaries does raise some interesting possibilities in terms of scattering mechanisms as such systems are known to be strongly effected by chirality and frequently explained in terms of pseudo-spin due to inequivalent electrons with Dirac dispersion relation.

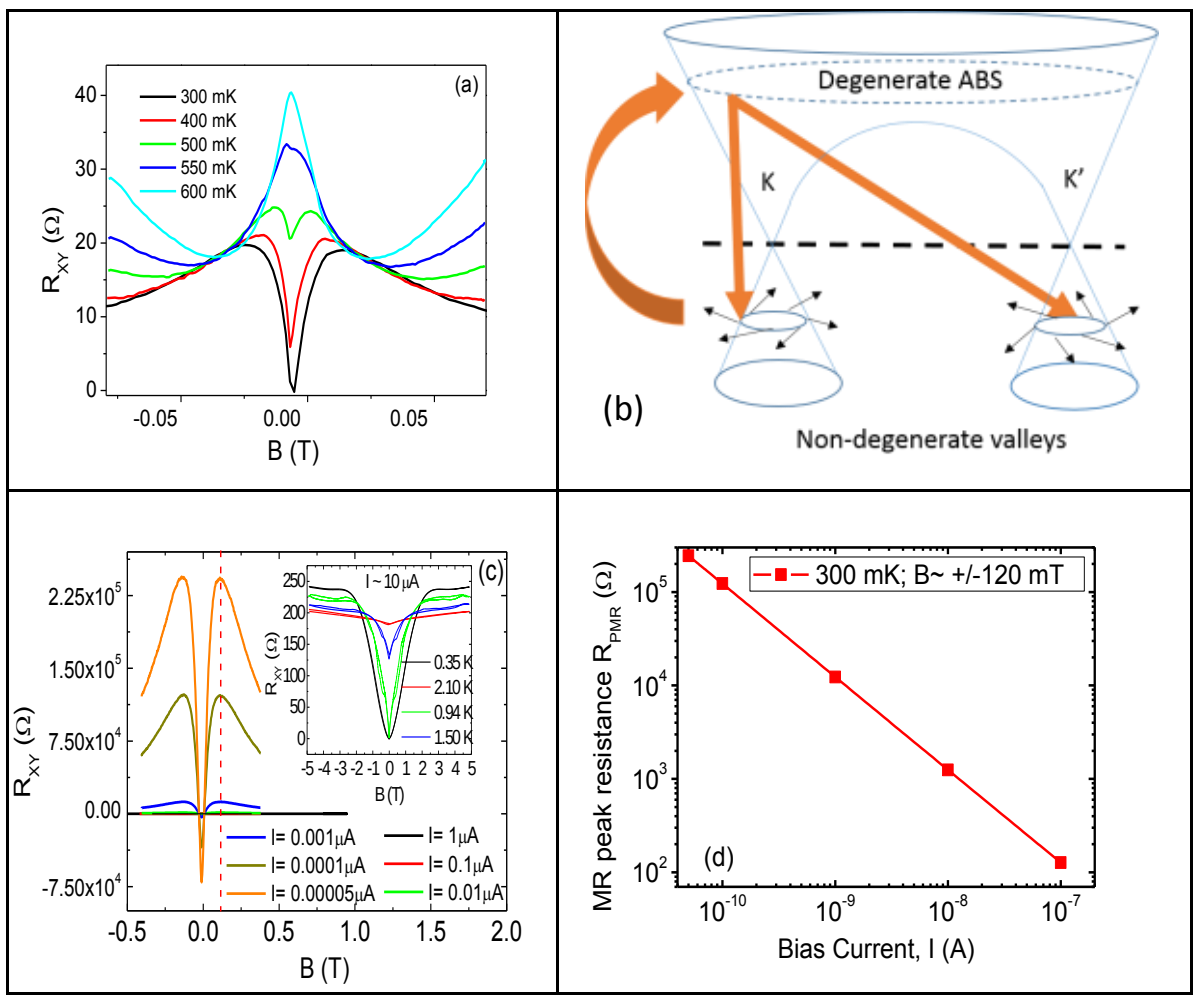

**Fig. 5** (a) Normalized magnetoconductance isotherms of sample B4, clearly showing the low field crossover from a negative (weak anti-localization) conductance regime to a positive (weak localization) conductance regime as temperature is increased, the critical temperature of this crossover was determined to be approximately that of the derived Kondo temperature for the sample. (b) The crossover from a WAL to a WL regime can be explained in terms of pseudo-spin coupling of chiral electron between valleys (inter) or within the same valley (intra), this is analogous to forming singlet and triplet states from pseudo-spin electrons, respectively. (c) The WAL features are also observed to be greatly enhanced by lowering measurement current. The inset indicates the thermal destruction of the WAL effect as temperature increases. (d) Similar to the behaviour of the resistance temperature data (Fig. 2b) the peak height decreases according to a power law as current is decreased, a possible indication that both features share the same origin.

One of the hallmark features of such pseudo-spin scattering effects is the crossover from WL to WAL with decreasing temperature, this is frequently observed in materials such as topological insulators as well as in graphene. This phenomenon has been explained in light of multi-valley nature of such materials and can be explained in terms of competition between bulk and surface states, as well as spin degeneracy lifting due to magnetic impurities in the surface states. As shown in figure 5 (a) when tuning the system to a state of finite resistance a definite crossover between WL and WAL as a function of temperature is indeed observed. As pointed out in previous sections, there is not much evidence suggesting spin-impurities exist in our system and that the transport properties are most likely to result from scattering relating to degenerate bound holes or ABS states, with this in mind it is possible to draw a comparison with the WAL to WL crossover observed in graphene where the transition takes place due to interference or scattering between chiral valleys (valley mixing) of different pseudo-spin. Essentially the scattering that takes place can be classified into two types; inter and intra-valley. In the first case electrons undergo scattering along the same (intra-) Dirac cone, this essentially means electron interference occurs between electrons of the same chirality, whereas in the second case electrons are scattered between different Dirac cones and hence scattering is dominated by electron interference between electrons of opposite chirality. The chirality of the electrons is manifest as a quantum degree of freedom, the pseudo-spin which must be taken into account when considering the pairing or scattering events. It is then possible for the scattering process to occur in four different ways, through the pairing of a singlet or one of three triplets states, the intra-valley scattering takes place due to the singlet and one triplet state where as the intervalley scattering occurs due to two triplet states.

With this description in mind we can explain the crossover in magnetoconductance observed in BNCD films which is most likely a result of an intermediate hole-electron interaction: At low temperatures Coulomb repulsion is enhanced, this opens a wider gap and leads to variation of band filling between the grains. At lower temperatures it is less likely for electrons to scatter between the grains (non-equivalent valleys), this case strongly favours intravalley (triplet state) scattering and WAL is observed. As the temperature is increased a regime where the electron-hole interaction (formation of the ABS) is reached, this state is degenerate and has a finite lifetime, depending on the state of the hole the electron binds to, upon de-excitation the electron will scatter to a valley of the same or opposite pseudo-spin with respect to its initial state, as is indicated in figure 5 (b). As the scattering process depends on degeneracy of the ABS it is analogous to a charge-Kondo effect described in previous sections. The de-excitation to inter or intra-Dirac cone is temperature dependent with intervalley scattering being favoured at higher temperatures, leading to (singlet) WL features in the magnetoresistance. Additionally, as indicated in figure 5 (c) it has been observed that the WAL effect peaks at a certain field after which negative magnetoresistance is observed with further increase of magnetic field. The peak magnetoresistance where the transition from a positive to a negative magnetoresistance regime occurs is seen to scale to a power law dependence with decreasing bias current, this is indicated in figure 5 (d). This trend has been observed in the temperature resistance dependent peaks as well and is thus believed to share the same origin i.e. the charge-Kondo effect. This again can be related to the symmetry of the order parameter as in the case of d-wave superconductors where low temperature thermodynamic as well as transport properties are described by power law temperature dependence instead of exponential dependence observed in isotropic order parameter s-wave materials [39].

## IV. CONCLUSIONS

In conclusion, we have investigated the occurrence of the so called bosonic insulator phase in nanocrystalline boron diamond films in terms of a charge-Kondo effect, we have established a $T^2$ dependence of the upturn of resistance and related the obtained Kondo temperature of each respective sample to the ratio of Coulomb charging to grain coupling, we find the experimental parameters follow the expected theoretical trend to some degree. Additionally we have investigated a Zero Bias Conductance peak which has not yet been reported for this materials. We find that the peak occurs above the critical temperature of each sample and scales with a power law relation with respect to temperature and applied field. The FWHM of the ZBCP follows the expected temperature dependence for Kondo system above the $T_c$ but however deviates drastically (again to a power law dependence) below $T_c$. We have also investigated the magnetic field behaviour of the ZBCP and found the system does not exhibit any Zeeman splitting, an indication of the presence of an Andreev Bound States instead of spin-impurities. The formation of the ABS via the acceptor hole is believed to restore degeneracy to the hole conduction channels thus allowing for the charge-Kondo effect. The magnetoresistance has also been investigated at lower bias current regime where a temperature dependent crossover from a weak anti-localization regime to a weak localization regime was observed, this transition is a Hallmark of topological conductance and was explained here in terms of valley mixing effects mediated due to chiral pseudo-spin scattering of the graphitic (graphene-like) boundaries. The overall current dependence shows that under certain conditions nanodiamond film behaves as an unconventional superconductor. The nature of the grain boundaries, comprising of graphene-like $sp^2$ hybridized carbon is expected to control many of the properties we observe in the boron doped superconducting system especially at low currents where the resistive features are more prominent and induce interesting features such as re-entrant behaviour as well as anomalous zero bias conductance. These transport features are clear indications of unconventional superconductivity in the system.

**Acknowledgements:** SB is grateful to Prof. Milos Nesladek (Hasselt University) for providing the diamond samples. SB would like to acknowledge the National Research Foundation, CSIR-NLC and the Wits URC for the financial support to conduct this research.
*Email: somnath.bhattacharyya@wits.ac.za